\documentstyle[12pt,iopconf]{article}
 \def\VEV#1{\left\langle #1\right\rangle}
\begin{document}

\title{Gravity's rainbow}
\author{George F. Smoot\dag\ and Paul J. Steinhardt\ddag}
\affil{\dag\ Lawrence Berkeley Laboratory, Space Sciences Laboratory \&
Center for Particle Astrophysics,
University of California, Berkeley, CA  94720}
\affil{\ddag\ Department of Physics, University of Pennsylvania,
Philadelphia, PA 19104}
\beginabstract
Could COBE DMR be detecting the imprint from a spectrum of gravitational waves
generated during inflation?  The conventional inflationary prediction
had been that
the cosmic microwave anisotropy is dominated by energy density fluctuations
generated during inflation and that
the gravitational waves contribute negligibly.
In this paper, we report on recent work that has shown that the conventional
wisdom may be wrong;  specifically, gravitational waves may dominate the
anisotropy in inflationary models where the spectrum of perturbations deviates
significantly from scale invariance (e.g., extended and power-law inflation
models and extreme versions of chaotic inflation).
If gravitational waves
do dominate  at the large-angular scales measured by COBE DMR, the expectation
and interpretation of anisotropies on small-angular
scales is profoundly altered.
\endabstract

\section{Introduction}
The recent COBE DMR (Smoot \etal 1992) observation of large-angular-scale
anisotropy in the cosmic microwave background (CMB) has opened a critical
empirical window for testing the inflationary model of the universe.
Ten years ago, inflation was  found to generate a nearly scale-invariant
(Harrison-Zel'dovich) spectrum of  energy density fluctuations
due to de Sitter fluctuations of  the inflaton field (Guth and Pi 1982,
Hawking 1982, Starobinskii 1982, Bardeen \etal 1983).  This  conclusion
led directly to
the conventional inflationary prediction
for how the CMB anisotropy varies on large-angular
scales (greater
than a few degrees) consistent with the COBE DMR measurements to date.
[At smaller-angular scales the original fluctuations have
re-entered the horizon prior to decoupling and have both evolved and been
augmented by other effects.
Consequently,
the prediction of the amplitude depends on
the amount and the nature (cold,  warm, hot) of the
dark matter.]

In recent months, though, a reconsideration of de Sitter fluctuations
and anisotropy
has shown that the conventional  prediction is  not generic
 (Davis \etal 1992).  For some inflationary models, there is a significant
 deviation from scale invariance, and the dominant fluctuations on
 COBE DMR scales is due to gravitational waves, rather than
 energy density fluctuations.
It has been well known for years (Starobinsky 1985; Abbott and Wise 1984)
that inflation produces  de Sitter
fluctuations of the  graviton (gravitational waves) as well as de Sitter
fluctuations of the inflaton.
 Whereas energy density fluctuations result in scalar fluctuations
of the metric, gravitational waves correspond to tensor perturbations.
Previously,
it had been thought that inflation predicted  that the
  tensor fluctuation amplitude is negligible compared to the
 scalar.  The recent reconsideration, though, has shown that this is only
 true for a restricted range of inflationary models.
 In cases where gravitational waves dominate on COBE DMR scales,
 extrapolation to smaller-angular
 scales and the interpretation for theories of large-scale structure are
 profoundly altered.

The  {\it Journ\'{e}es Relativistes 1992} acted as a catalyst for
initiating the re-examination  of  the gravitational waves from inflation.
the Meeting
took place at a fortuitous time, less than a month after the  dramatic
COBE DMR announcement  had focused international attention on
 CMB anisotropies.  One of us (GFS) had come to Amsterdam with a
  crude argument which suggested that tensor and scalar fluctuations
  from inflation should be comparable in amplitude and wanted to see
  if a more rigorous argument could be found (see  Section~2).
  Along the beautiful canals of Amsterdam, a discussion began
  between the two authors which ultimately culminated in an
 intense collaboration with
  Rick Davis, Hardy
  Hodges, and Mike Turner
  (Davis \etal\ 1992)
  to resolve the issue.
  (See Acknowledgements for a list of other authors who have  also
  considered this issue.)
  In  appreciation for the special role that
  the {\it Journ\'{e}es Relativistes}, Amsterdam, and our gracious hosts
  played in this effort,
  we would like to present this work, including  pedagogical points and
  some additional results that could not be included in the original
  paper.

  Section~2 is intended as a heuristic introduction to  the  inflationary
  fluctuation spectrum.  The remaining sections are more technical,
  giving precise details on  how to compute
  the anisotropy in the CMB due to inflaton and graviton fluctuations
  on large-angular scales using the Sachs-Wolfe effect.
  Section~3  focuses on the scalar (energy density) perturbations and
  Section~4 focuses on the tensor (gravitational wave) perturbations.
  Section~5 discusses the tensor-to-scalar ratio and includes
  a  model-independent expression relating this  ratio to the
  tilt of the fluctuation spectrum away from scale-invariance.
   Section~6   describes
  the predictions  for specific  types of inflationary models.
  Section~7 discusses the implications for COBE DMR, for experiments
  at smaller-angular scales, and for models of large-scale structure.

\section{General and Pedagogical Discussion}

Why do we say that the recent COBE DMR (Smoot \etal 1992) observation
of large-angular-scale anisotropy in the cosmic microwave background (CMB)
show there is a chance to test inflationary models?
COBE DMR measures directly the angular variation in CMB temperature near
the earth but we interpret it as measurements of the curvature fluctuations
at the surface of last scattering, probably at a red shift of about $10^3$.
We then use that information to distinguish various scenarios describing
what was going on in the Universe in the first fraction of a second
($10^{-43} s < t \leq 10^{-9}s$), since different scenarios make
different predictions about the spectrum of curvature fluctuations.
This section presents a heuristic, pedagogical
discussion of how this works and what
we are testing.

\subsection{Curvature Fluctuations and Temperature Anisotropy}

To understand the effect of curvature perturbations at the surface of
last scattering,  we must propagate the photons from their last interaction
to the present epoch observer.
The specific number density of photons is
\begin{equation}
n_{\gamma} = h^{-4}  \frac{I(\nu)}{\nu^3} \;
\end{equation}
where $h$ is Planck's constant, $I(\nu)$ is the specific intensity,
and $\nu$ is the frequency.
Since during their propagation the number of photons is conserved,
($I(\nu)/\nu^3$) is a conserved quantity.
Likewise for a blackbody spectrum ($T/\nu$) is a conserved quantity.
Whatever causes a frequency shift in the photons also causes a
compensating aberration of solid angle and relevant quantities
just so as to maintain a blackbody form for the spectrum.  The
only change is that the temperature in the Planck formula
is replaced by a red shifted value.

Curvature fluctuations perturb the CMB temperature because they
lead to variations of the gravitational potential on the surface
of last scattering.  To see how gravitational potential variations
affect the CMB temperature,
consider a simple analogy, the cosmic gravitational red shift experiment.
The standard gravitational red shift,
sometimes called gravitational time dilation,
can be understood either through the equivalence principle or time dilation.
A third view is a little more heuristic, namely conservation of energy.
Quantum mechanics tells us that a photon of frequency $\nu$ has energy
$E~=~h\nu$, where h is Planck's constant.
When a photon changes its location in a static potential field, its frequency
must change in an amount that corresponds to the energy change.
The energy change is
$ h \Delta \nu \approx E/c^2 ~ \Delta \phi ~=~ h\nu \Delta \phi /c^2 $
so that
\begin{equation}
\frac{\Delta \nu}{\nu} \; = \frac{\Delta \phi}{c^2} \; .
\end{equation}
%
%
Then, if two  emitters emit 
blackbody radiation
at temperature $T_e$ from two points with differing
gravitational potential, the observer sees a temperature variation of
\begin{equation} \label{eq4}
\frac{\delta T}{T} \; = \frac{\delta \phi}{c^2} \; .
\end{equation}

In the cosmological problem of interest, Eq.~(\ref{eq4}) must be
amended because the universe is expanding and density perturbations are
growing. The growth of the density perturbation partially compensates
for the gravitational red shift effect. In a matter-dominated expanding
universe
perturbations grow linearly with the scale factor $a(t) \propto t^{2/3}$,
so that there is an additional contribution to the anisotropy
$\propto \delta a/a$, which is also
proportional to $\Delta \phi$.   Re-expressing this as a fluctuation in
the gravitational potential (using the Newtonian approximation to
the metric), one finds that a net anisotropy:
\begin{equation}
\frac{\delta T}{T} \; = \frac{1}{3} \; \frac{\delta \phi}{c^2} \;
\end{equation}

In 1967 Sachs and Wolfe derived  the formal first-order
formula for calculating
the effect of metric fluctuations in
the CMB temperature which included both the variation in the temperature
on the surface of last scattering and the induced variation
in the expansion rate.
Their results were
\begin{equation}
T_o = T_e (\frac{\tau_e}{\tau_o} \;)^2 (1 + \frac{\delta T}{T} )
= T_e \frac{a(t_e)}{a(t_r)} \; (1 + \frac{\delta T}{T} )
= \frac{T_e}{1+z} \; (1 + \frac{\delta T}{T} )
\end{equation}
where $T_o$ and $T_e$ are the temperature at the observer and emitter
respectively, $\tau_e$ and $\tau_o$ are the conformal times of emission
and reception, approximately 1/30 and 1 respectively,
and
\begin{equation} \label{SW}
\delta T / T = \frac{1}{2} \;
\int_{\tau_e}^{\tau_o} ( \frac{d h_{\mu\nu}}{d\tau} e^{\mu} e^{\nu}
- 2 \frac{ \partial h_{0 \nu}}{\partial \tau} e^{\nu} ) d\lambda
\end{equation}
where ${h}_{\mu \nu}$ is the metric tensor fluctuation,
$e^{\mu}$ is the direction vector,
and $\tau$ is the conformal time ($d\tau = dt/a(t)$).
Their work showed that the variation in expansion rate due
to   potential variation
along a  surface  of uniform temperature  is approximately
%
%
\begin{equation} \label{eq5}
\delta T / T \approx \frac{1}{3} \; \delta \phi / c^2
= - \frac{1}{3} \; \delta \rho / \rho_c (d/d_H)^2
\end{equation}
where $\delta \phi$ is potential variation at the surface of last scattering,
$\delta \rho$ is the density variation producing the potential difference,
$\rho_c$ is the critical density and $d/d_H$ is the ratio of the scale
size of the fluctuation compared the the horizon size at last scattering.
Note that
the potential variations on the surface of last scattering
are more significant than time varying potentials along the photon's path
 (whose net effect averages to zero).

Finally, Eq.~(\ref{eq5}) assumes that the
emission from the various locations on the surface of last scattering
is from the same temperature blackbody. Density variations
along the surface of last scattering will produce variations in the number
density of photons.
For adiabatic fluctuations
$ \delta n_{\gamma}/{n_{\gamma}} ~=~ \delta \rho / \rho$
where $\rho$ is the density,
which results in a local temperature variation
$\delta T/ T ~=~ 1/3 ~ \delta \rho / \rho$.  Hence,
there are, in principle,
two competing effects with opposite signs for an adiabatic/curvature
fluctuation: (1) the temperature increase due to the greater photon number
density
and (2) the temperature decrease due to gravitational red shift of the
higher gravitational potential caused by the fluctuation
\begin{equation}
\delta T / T = \frac{1}{3} \; \frac{\delta \rho}{\rho} ~
[ 1 -  \frac{\rho}{\rho_c} (\frac{d}{d_H})^2 ]
\end{equation}
Which term dominates depends upon the angular (or physical) scale of interest.
For fluctuations equal to or greater than the horizon size at the
surface of last scattering, the Sachs-Wolfe effect (Eq.~(\ref{eq5}))
dominates.
For fluctuations equal to or smaller than the horizon size at the surface
of last scattering, the adiabatic temperature fluctuation would dominate
over the Sachs-Wolfe effect. However, other evolutionary
effects become even more important at small angles
including the  Doppler effect from the gravitational
collapse of density perturbations or Thomson scattering if the universe
is reionized.

For the purposes of this paper, though,  we will consider only
the large-angular scales measured by COBE DMR for which
only the perturbation given  by    Eq.~(\ref{eq5}) is significant.
The horizon size if the surface of last
scattering at the canonical red shift of 10$^3$ is about 1 degree.
If the universe has been re-ionized, then the horizon size might be as large
as a few degrees.

\subsection{Inflation and the Origin of Fluctuations}

Models of inflation provide an explanation of the mechanism for driving
the expansion through a near de Sitter phase and for generating
a spectrum of
scalar and tensor metric fluctuations on cosmological scales.
In the most simple models,  the  de Sitter expansion comes about
as a scalar field evolves slowly from a state of high vacuum  energy
to a state of low   vacuum energy (see the article by Steinhardt in
this volume). The high energy of
the initial state dominates the energy density of the universe
 and acts as a large, positive cosmological constant.

The equation of motion for the scalar field during de Sitter expansion is
\begin{equation}
d^2 \phi / dt^2 + 3 H d \phi / dt + V'(\phi)  = 0
\end{equation}
where $\phi$ is the scalar field strength, $V'(\phi)$ is the
scalar field potential gradient at $\phi$, and the $3H d\phi / dt$ term
is the damping or energy loss term coming from the expansion
of the universe. This  damping
is the cosmological red shift of the momentum in the scalar field.
In the slow roll condition when the gradient of the scale
is not large $d^2 \phi / dt^2 $ is negligible and
\begin{equation} \label{slow-roll}
3 H d \phi / dt \approx -V'(\phi).
\end{equation}

Quantum mechanical fluctuations in the scalar field, $\delta \phi$,
will speed up or slow down the local expansion rate compared to the
global rate and thus produce curvature fluctuations.
Quantum fluctuations in the graviton field, $\delta \phi_{GW}$,
generate gravitational waves, tensor metric perturbations. 
The basic source of fluctuations is the same for both:
zero point quantum fluctuations or the equivalent de Sitter curvature
thermal radiation at the Gibbons-Hawking temperature $(T_{GH} = H/2\pi)$.

Although  a rigorous calculation entails many subtleties, especially
 the hypersurface- or gauge-dependence of the fluctuation amplitude
 when it is inflated outside the horizon, there is a useful
 mnemonic for the final, correct answer.  First, the scalar
curvature fluctuations on a given scale as that scale
is stretched outside the horizon during the de Sitter (inflationary)
epoch
 can be understood as a local variation
in the universe's scale factor $a$.
\begin{equation} \label{eqs}
\delta {\rm ln} a = \frac{d {\rm ln} a}{d \phi}  \delta \phi
= \frac{d {\rm ln} a}{dt} / \frac{d \phi}{dt} ~ \delta \phi
= H / \frac{d \phi}{dt} ~ \delta \phi ;
{}~~~~~~ H \equiv \frac{d {\rm ln} a}{dt}.
\end{equation}
In a de Sitter background,
the rms fluctuation in the inflaton is $\VEV{\delta \phi}^{1/2} = H/2 \pi$.
Tensor fluctuations in the metric  are produced by the same
basic process  and in rough equipartition
but with a coupling of of $\sqrt{16\pi G}$:
\begin{equation} \label{eqg}
 h_{GW} = \sqrt{16\pi G} \delta \phi_{GW} = \sqrt{16\pi} \delta \phi_{GW}
/m_{Pl},
\end{equation}
where each component of the metric undergoes fluctuations with rms
amplitude $\VEV{\delta \phi_{GW}}^{1/2}= H/2 \pi$.
(One must include
the fact that there are two tensor degrees of freedom but only one scalar.)

How do these curvature fluctuations evolve to the surface of last scattering?
For both scalar and  tensor fluctuations, the amplitude for a
given wavelength is frozen  from when the wavelength is stretched
outside the horizon during the de Sitter epoch to when it re-enters
the horizon during the Friedmann-Robertson-Walker epoch that follows.
 Hence, the
amplitude upon re-entering the horizon simply equals the amplitude
 when exiting during the de Sitter epoch.
 Hence,
for gravity waves, the amplitude  on the surface of last scattering is
determined from Eq.~(\ref{eqg}),
$\VEV{h_{GW}^2}^{1/2} \sim (2/\pi)^{1/2} H/m_{Pl}$.
For the scalar field fluctuations, the curvature perturbation
$\delta \; {\rm ln} \; a$ during the de Sitter epoch
 can be  related to
  the fluctuations in the density
through the relativistic continuity equation
\begin{equation}
\frac{d \rho}{dt} + 3 H (\rho + p/c^2) = 0 ;   ~~~~~ H \equiv \frac{d
ln(a)}{dt},
\end{equation}
where $p$ is the pressure.
Multiplying through by $\delta t$ 
we have
\begin{equation}
\delta \rho + 3(\rho + p/c^2) \delta ln(a) = 0
{}~~ \rm{or} ~~
\delta ln(a) = - \frac{1}{3} \frac{\delta \rho}{(\rho + p/c^2)}.
\end{equation}
(The last expression  has been shown to be
  gauge-invariant (Bardeen \etal 1983).)
The curvature perturbation when a given scale re-enters the
horizon in the Friedmann-Robertson-Walker epoch  equals
$\delta ln(a) = - \frac{1}{3} \frac{\delta \rho}{(\rho + p/c^2)}$
when the scale was stretched beyond the horizon during the
de Sitter epoch.
{}From the slow roll condition, Eq.~(\ref{slow-roll}),
and using  $H^2 ~=~ 8\pi V(\phi)/3 m_{Pl}^2$,
one  finds
\begin{equation}
\delta {\rm ln} a = -\frac{3H^2}{V'(\phi)} ~ \delta \phi
= -\frac{8\pi V}{V'(\phi) m_{Pl}^2} ~ \delta \phi .
\end{equation}
Thus the scalar (density) metric fluctuations amplitude
depends not only on the energy in the scalar field
(on the expansion rate)
but also on the gradient of the field's potential.

In sum, we estimate that
the ratio of rms gravity wave curvature fluctuations to those from the
scalar field is
\begin{equation}
\frac{< h_{GW}^2 >^{1/2}}{< h_{\phi}^2 >^{1/2}}
= \sqrt{ \frac{2}{\pi}}  ~ \frac{V'(\phi) m_{Pl}}{V},
\end{equation}
where this expression is to be evaluated when the scale of
interest was stretched beyond the horizon during the de Sitter
epoch.  For COBE DMR scales, this corresponds to roughly
60 e-foldings before the end of inflation.
This estimate  is in good agreement with the rigorous calculations
in later sections of this paper.

In joining together to collaborate on this problem,
one of us (GFS) argued that equipartition implies that
the ratio of tensor to scalar  fluctuations must be comparable.
The other (PJS)  was convinced that  the tensor fluctuations
from inflation
must be negligible based on the experience  from well-established calculations
(Abbott and Wise 1984b)
for the earliest  models
of inflation.
  In fact, neither was correct.
 The  equipartition argument is deceiving because
 the ratio of tensor to scalar contributions
 depends on conditions during the last 60 e-foldings of
 inflation, which is a period when the  expansion is beginning to
 decrease and the de Sitter approximation is becoming invalid.
 The experience from earlier models of inflation is deceiving
 because those models entailed very flat potentials, whereas other
 inflationary model  use steeper potentials.
  As we have seen from the heuristic argument above, the
  ratio of tensor to scalar contributions is proportional
  to $V'$.  A  rigorous derivation is given in the next few
  sections.
 In re-examining more general models of inflation, we found that   the
 precise rate at which the universe exits inflation differs among
 inflationary models,  as does $V'$,
 and,  surprisingly, some inflationary models predict
 a tensor contribution that dominates on COBE DMR scales.

\section{Scalar metric fluctuations from inflation}

The fluctuations of the metric can  be divided into scalar and tensor modes.
Choosing coordinates such that small metric fluctuations $h_{\mu\nu}$
satisfy $h_{00}=h_{0i}=h_{i0}=0$, the remaining non-zero components can
be written:
\begin{equation}
h_{ij} = \frac{1}{3} \; h \delta_{ij} + \tilde{h}_{ij}
\end{equation}
where $h$ is the scalar mode and  $\tilde{h}_{ij}$ is the (traceless) tensor
mode.

In Section~2.2 (and in the  contribution by Steinhardt in this volume),
we have described the primordial spectrum of scalar and tensor fluctuations
of the metric that are generated in inflationary models.  The spectra
are determined by the inflaton potential $V$ and its slope $V'$.  Beginning
with this section, we compute how these metric fluctuations leave their
imprint on the CMB anisotropy through the Sachs-Wolfe effect
(Sachs and Wolfe 1967).

The temperature fluctuations can be decomposed into spherical-harmonic
amplitudes:
\begin{equation}
\Delta T/T = \sum_{\ell, m} a_{\ell m} Y_{\ell m}(\theta, \phi).
\end{equation}
The angular temperature autocorrelation function can be written
(Peebles 1982):
\begin{equation}
C(\alpha) = \VEV{\frac{\Delta T}{T}({\bf q}) \frac{\Delta T}{T}({\bf q}')} =
\frac{1}{4 \pi} \sum_{\ell} (2 \ell +1) \VEV{a_{\ell}^2} P_{\ell}
(\cos{\alpha})
\end{equation}
where  ${\bf q} \cdot {\bf q}' = \; \cos{\alpha}$ and
\begin{equation}
\VEV{a_{\ell}^2} \equiv \VEV{\sum_{m=-\ell}^{m=\ell}  |a_{\ell m}|^2}.
\end{equation}

For a  gaussian distribution of fluctuations, the
multipole moments computed from the Sachs-Wolfe
expression, Eq.~(\ref{SW}), are (Abbott and Wise 1984a):
\begin{equation}\label{scalint}
\VEV{a_{\ell}^2} = \frac{4 \pi^2}{2 \ell +1}
\int_0^{\omega_{max}} \frac{d \omega}{\omega}  \epsilon_H^2 (\omega)
\left\{ (2 \ell+1) j_{\ell}(\omega) +\frac{\omega\; \tau_{R}}{\tau_0 - \tau_R}
[\ell j_{\ell-1}(\omega)- (\ell+1) j_{\ell+1}(\omega)]\right\}^2,
\end{equation}
 where   $\tau_0$ is  present conformal time, $\tau_R$ is conformal time
 at recombination , $\omega\equiv k( \tau_0-\tau_R)$
 labels the wavevector of the of
 the Fourier mode, $ \omega_{max} $ is set by
 modes comparable to the decoupling horizon, and
  $\epsilon_H (\omega)$ is the
  gauge-invariant fluctuation amplitude (Bardeen \etal
 1983). (Note that we choose units where $k \tau /2 =1$ corresponds to
 horizon crossing.)

The amplitude $\epsilon_H$
is related to the power
spectrum by (Abbott and Schaefer 1986)
\begin{equation}
\epsilon_H^2 = \frac{k^3 \delta_k^2}{(2 \pi)^3},
\end{equation}
where
 the power spectrum from inflation is what was heuristically derived in
 Section~2.2 [see Eq.~(\ref{eqs})]:
\begin{equation}
\frac{k^\frac{3}{2} \delta_k}{\sqrt{2} \pi} = \frac{H^2}
{5 \pi\dot{\phi}}|_{N\sim 60}.
\end{equation}
 Recall that $\phi$ is the scalar inflaton, $H$ is the Hubble
parameter, and the rhs is to be evaluated $N \sim 60$ e-foldings before the end
of
inflation when fluctuations on CMB length scales crossed outside the
horizon (Bardeen \etal 1983).  Consequently,
$\epsilon_H = \frac{H^2}
{10 \pi^{3/2}\dot{\phi}}|_{N\sim 60}$.
Combining these relations, we find
a quadrupole moment for the scalar ($S$) fluctuations:
\begin{equation} \label{scalar}
S \equiv  \VEV{a_2^2}_S \equiv \VEV{\sum_{m=-2}^{m=2}|a_{2m}|^2}
 =  \frac{1}{60 \pi} \frac{H^4}{\dot{\phi}^2}
  \end{equation}
  From Eq.~(\ref{scalint}), the high multipole moments can be determined
  from the quadrupole.  Finally, Eq.~(\ref{scalar}) can be re-expressed
  in terms of the inflaton potential, $V(\phi)$ using the
  equation of motion, Eq.~(\ref{slow-roll}); for a strictly
  scale-invariant spectrum, we find:
\begin{equation} \label{squad}
S \equiv 28.1 \frac{V^3}{V'^2 m_{Pl}^6},
\end{equation}
where  $m_{Pl} = 1.22 \times 10^{19}$~GeV is the Planck mass, and we
have used the Einstein equation, $ H^2=8 \pi V/3m_{Pl}^2$,
and the slow-roll condition, $3 H \dot{\phi} = -  V'$.
[Strict scale invariance is not achievable in reality because
there is always some change in the slope $V'$ near the end of
inflation; however,  Eq.~(\ref{squad}) is a useful first
approximation.]

\section{Tensor metric fluctuations from inflation}
De Sitter fluctuations of the massless excitations of the metric  induce
gravitational waves.  The gravitational waves also  generate CMB anisotropy
on large-angular scales through the Sachs-Wolfe effect (Abbott and
Wise 1984b).  The result can
be expressed in an integral formula analogous to  Eq.~(\ref{scalint}):
\begin{equation} \label{tenint}
\VEV{a_{\ell}^2} = 72 \pi^2 \ell (\ell-1)(\ell+1)(\ell+2)(2 \ell+1)
\int_{0}^{\omega_{max}} \omega d\omega \left\{\int_0^{1 - (\tau_R/\tau_0)}d z
A^2(k) j_1'(\omega(1-z))
I[\omega z] \right\}^2
\end{equation}
where
\begin{equation}
I[\omega z]=
\frac{2 j_{\ell}(\omega z)}{(2 \ell-1) (2 \ell+3)}
+\frac{ j_{\ell-2}(\omega z)}{(2 \ell-1) (2 \ell+1)}
+\frac{ j_{\ell+2}(\omega z)}{(2 \ell+1) (2 \ell+3 )}.
\end{equation}
Here, the tensor mode power spectrum $A^2(k)= V/m_{Pl}^4$ is constant for a
strictly scale-invariant spectrum.  Evaluating this expression, we
find  for the tensor quadrupole moment for a strictly scale-invariant spectrum:
\begin{equation} \label{tquad}
 T \equiv 7.7 \;  \frac{V}{m_{Pl}^4}
 \end{equation}
 [Once again, strict scale invariance (constant $V$) is not achievable
 in any realistic model, but this   is a good first approximation.]

\section{Central result: the ratio of tensor to scalar modes}
{}From Eqs.~(\ref{squad}) and~(\ref{tquad}), we obtain the ratio of tensor
to scalar quadrupole anisotropies for a scale-invariant spectrum:
\begin{equation} \label{ratio}
\frac{T}{S} \equiv \frac{\VEV{a_2^2}_T }{ \VEV{a_2^2}_S } \approx
0.28
\left(\frac{ V' m_{Pl}}{V}
\right)^2|_{N \sim 60},
\end{equation}

As we have noted,
the coefficients in Eqs.~(\ref{squad}) and~(\ref{tquad})
were derived assuming strict scale invariance (e.g, constant $\epsilon_H$
and $A^2$).
This limit
corresponds to  constant $V$, a limit which cannot be achieved exactly
in inflationary models
if the inflaton is to slow-roll from the true to false vacuum
phase.
Hence, we will  need
to compute $T/S$ for spectra that are  ``tilted" away from strict
scale-invariance, $\epsilon_H^2$ and $A^2$ proportional to
$k^{(1-n)}$ (using the usual convention which identifies $n=1$
as the spectral index for a strictly scale-invariant spectrum).
As  Table I below shows, tilt can  significantly affect the
coefficients  in the expressions for $T$ and $S$ individually, but
the ratio
 remains nearly constant even for large tilts:
\begin{center}
 \begin{tabular}{cccc}
 \bottomline
 \\ Tilt & $T$ Coeff. & $S$ Coeff. & $T/S$ Coeff. \\
 (n) & (Eq.~\ref{tquad}) & (Eq.~\ref{squad}) &(Eq.~\ref{ratio}) \\
 \hline \hline \\
 1 & 7.7 &  29 & 0.28 \\
 0.85 & 6.1 & 25 & 0.25 \\
 0.75 & 5.4 & 21  & 0.25 \\
 0.50 & 4.0 & 17 & 0.24 \\
 0.0 & 2.3 & 10 & 0.22  \\
 \topline
 \end{tabular}
\end{center}

\vspace*{.1in}

Although the ratio of the dimensionless coefficients is roughly independent of
tilt, the ratio of the dimensionful parameters in Eqs.~(\ref{squad})
and~(\ref{tquad}) are  highly sensitive tilt.  Consequently, there is
a direct  relation between $T/S$ and $n$ that can be derived in a
model-independent way:
The ratio of tensor to scalar perturbations is controlled by
the steepness of the potential, $V^\prime m_{Pl}/V$; cf.~Eq.~(\ref{ratio}).
During inflation, this quantity also determines the
ratio of the kinetic to potential energy of the
scalar field (Steinhardt and Turner 1984), $\frac{1}{2}{\dot\phi}^2/V
\simeq (V^\prime m_{Pl} /V)^2/48\pi$, which in turn determines the
effective equation of state ($p=\gamma \rho$) and
the evolution of the cosmic-scale factor ($R\propto t^m$):
$\gamma= [\frac{1}{2}{\dot\phi}^2 -V]/[\frac{1}{2}{\dot\phi}^2
+V]$ and $m= 2/3(1+\gamma )$ (during inflation
$\gamma$ and $m$ can vary).  It is simple to show that
the tensor perturbations are characterized
by a power spectrum $|\delta^T_k|^2\propto k^{n_{T}-1}$
and the scalar (density) perturbations
by $|\delta^S_k|^2 \propto k^n$,
where $n_T=(m-3)/(m-1)$.

Note that we have assigned different
symbols for the scalar
and tensor exponents (power indices). The indices are determined
by how much the expressions on the rhs of Eqs.~(\ref{squad}) and~(\ref{tquad})
change  near the last 60 e-foldings of inflation.
 The key
difference is that the tensor  contribution  in Eq.~(\ref{tquad}) depends
only on how $V$ changes in the last 60 e-foldings, whereas the
scalar contribution depends on $V$ and $V'$.
For most inflation models,
$n$ and $n_T$ are
identical to within a few percent.
 However, it is possible to  design potentials such that
 the two power indices are different.  For example,  if
 $\phi$ rolls past an inflection point ($V'=0$)
 within the last 60 e-folds of inflation, this dramatically changes the
 scalar power spectrum but does not affect the tensor
 power spectrum
 (see the contribution by PJS in this volume for an example).
In fact, $n_T =
(m-3)/(m-1)$ is strictly less than unity, but $n$  can
exceed unity over COBE scales.
Having issued this warning, we will now restrict the discussion to
more typical models where the distinction between $n$ and $n_T$
is negligible.

 The expansion-rate index $m$ and the power-spectrum
 index $n$ (for $N\sim 60$)  can be directly tied to   $T/S$:
 \begin{equation}  \label{mn}
 m =  14\left(\frac{S}{T}\right) + 1/3 \simeq 14\left(\frac{S}{T}\right) ;
 \qquad  n = 1 - {3 \; (T/S) \over 21- \; (T/S)} \simeq 1 -
 \frac{1}{7}\left( \frac{T}{S}\right).
 \end{equation}
  If the tensor mode is
 to dominate---i.e., $T/S\ge 1$---then
 $m$ must be less than 14 and $n$
 must be less than  0.85.
 (Conversely,
 in models where the expansion is exponential
 and the spectrum is scale invariant, the ratio of tensor to
 scalar is very small.)

 From the fact that inflation must
 be ``superluminal'' ($m > 1$), we can use Eq.~(\ref{mn}) to
 derive an approximate {\it upper}
 bound, $T/S \le 20$.  However,
 the COBE DMR  bound on the power-spectrum index $n$,
 $n=1.1\pm 0.6$, which implies that $n \ge 0.5$ when
 $T/S\ge 1$, leads to the stronger limit, $T/S\le 3$ (and $m\ge 5$).
 (There are yet stronger bounds on $n$ based
 upon structure formation).

\section{Predictions of inflationary models}
The first workable  inflationary model, ``new inflation,'' is characterized
by a flat potential in which $V$ is nearly constant during inflation
(Linde 1982; Albrecht and Steinhardt 1982).
As noted in the section above, the fluctuation spectrum has nearly
zero tilt in this limit.  Also, since $V' \rightarrow 0$  in
this limit, the ratio $T/S \rightarrow 0$ and, hence, the tensor
contribution to the CMB anisotropy is negligible.  It is from this
example that the conventional  inflationary prediction --- the
scalar mode dominates the tensor mode --- has been drawn.
However,  alternative models of inflation have been developed to
resolve the fine-tuning problem in new inflation.  Some of these
alternative models  lead to non-negligible tilt
 (see, for example, the contribution by PJS in this volume) and to
 non-negligible tensor contribution to the CMB anisotropy. We
 will outline below the analysis of a few of  these models:

{\it Extended   and power-law  inflation
models} can be described in terms of a potential of the form,
$V(\phi) = V_0 \; {\exp} (-\beta \phi/m_{Pl} )$,
where $\beta$ is constant or slowly time-dependent.
In extended inflation (La and Steinhardt 1989)
$\phi$ is related to a field
that is coupled to the scalar curvature
(e.g., a dilaton or Brans-Dicke field), which leads to a modification
of Einstein gravity.  The modified gravity action
can be re-expressed via a Weyl transformation
as the usual Einstein action plus a minimally coupled scalar
field ($\phi$) with an exponential potential.  In the simplest
example of extended inflation (La and Steinhardt 1989), $\beta =
\sqrt{64 \pi/(2 \omega +3)}$, where
$\omega$ is the Brans-Dicke parameter.
For an exponential potential, Eq.~(\ref{ratio}) implies:
\begin{equation}  \label{exp}
{T\over S} \approx 0.28 \beta^2 =  \frac{56}{ 2 \omega +3}
\end{equation}
The ratio $T/S \ge 1$ for
  $\omega \le 26$ ($\beta \ge 1.9$).  Interestingly,
  $\omega \le 26$ is almost precisely what is required to avoid
  unacceptable inhomogeneities  from  big bubbles
  in extended inflation (Weinberg 1989; La \etal 1989).
  (Though $\omega\le 26$ is inconsistent
  with solar-system limits for Brans-Dicke theory,
  these constraints are evaded by giving the Brans-Dicke field
  a mass.) The main distinction between extended inflation and
  power-law inflation (Lucchin and Matarrese 1985) is the mechanism
  for ending inflation.  In extended inflation,  superluminal
  expansion  is driven by an ordinary first order phase
  transition that ends by bubble nucleation without fine-tuning of
  parameters.
  Power-law inflation ends by slow-roll down the exponential potential to
  a stable vacuum phase.  Reheating occurs through oscillation about
  a steep minimum and
  decay of the inflaton field, which requires  jerryrigging the
  potential so that the exponential potential  bottoms out into a
  steep harmonic minimum.  It seems fair to say that, although
  both lead to power-law expansion rates,  extended inflation is
  significantly more natural.

  {\it Chaotic inflation models }
  typically invoke a potential of the form, $V(\phi) = \lambda \, \phi^p$,
  where $\phi  \gg m_{Pl}$ initially, and rolls to $\phi =0$ (Linde 1983). The
  ratio of tensor to scalar anisotropies can be
  expressed in terms of $\phi_N$, the value of the scalar field
  $N \sim 60$ e-foldings before the end of inflation.  Using the relation,
  \begin{equation}  \label{Nchaos}
  N(\phi ) = \int_{t_{\rm end}}^{t_N} H dt =\frac{8 \pi}{m_{Pl}^2}
   \int_{\phi_{\rm end}}^{\phi_N} \frac{V}{V'} d \phi =
    \frac{4\pi}{p}\frac{\phi^2}{m_{Pl}^2} - \frac{p}{12},
    \end{equation}
    where $\phi_{\rm end}^2=p^2m_{Pl}^2/48\pi$,
  one finds that:
    \begin{equation} \label{chaos}
    \frac{T}{S} \approx \frac{p}{17.4}\left[ 1 + \frac{p}{720}\right]^{-1},
    \end{equation}
    where we have set $N=60$. (A similar result was obtained
    earlier by Starobinskii (1985).) For the chaotic-inflation
    models usually discussed, $p=2$ and $4$,
    the scalar mode dominates:  $T/S=0.11$ and 0.23;
    however, for extreme models, $p\ge 18$, the tensor
    mode could dominate.

	We have examined a variety of other models, e.g., cosine and
	polynomial potentials; the only other example that we found
	that permits $T/S \ge 1$ is $V(\phi)= \lambda (\phi^2 - \sigma^2)^2$.
	For $\sigma \le 0.8 \; m_{Pl}$, $T/S \ge 1$. However, $\sigma \ge
	0.5 \; m_{Pl}$ is required to obtain sufficient inflation; hence,
	this example is marginal in that $T/S \ge 1$ for only a narrow
	range of allowed parameters.

\section{Implications if the tensor mode dominates COBE DMR measurements}

COBE DMR alone is not able to distinguish  whether the scalar or tensor
dominates on large-angular scales.  First, the measured bound on the
power index, $n= 1.1 \pm 0.6$ (one-$\sigma$ error) is so broad that
it embraces models
ranging from scale-invariant ($n=1$ --- for which the scalar mode dominates)
to  $n<0.8$, for which the tensor mode dominates.
Secondly, the shape of the temperature autocorrelation function on
large-angular scales varies only weakly between tensor and scalar
type fluctuations.
In Figure 1, we
have shown the temperature autocorrelation function  predicted for
a scale-invariant spectrum of pure scalar vs. pure tensor mode, where
both have been normalized to  the same quadrupole moment.
As in the COBE DMR, we have subtracted the
monopole through quadrupole contributions.  Superimposed
are the observational error  and the theoretical error due to
``cosmic variance."   (Inflation predicts the statistical properties
of the CMB anisotropy averaged over all observation points in the
universe; since  the observations are made from only one vantage
point, this introduces a theoretical error in the prediction.)

\begin{figure}
\vspace*{6in}
\caption{
Temperature autocorrelation function (from the Sachs-Wolfe
effect) for tensor and scalar modes each normalized to the COBE DMR
quadrupole anisotropy using a scale-invariant ($n=1$) spectrum and
the COBE DMR window function (Smoot \etal 1992).
Tensor and scalar modes are distinguishable at small angles, but
COBE  DMR, whose data is superimposed in (a), is unable to
resolve the difference.  Cosmic variance, shown in (b), is a
theoretical uncertainty that adds to the inability to distinguish.
CDM predictions (Holtzman 1989) for the scalar contribution to
$C(0)$ (assuming $h=0.5$ and $\Omega_b=0.1$) is shown for bias
factors $b=1$ and 1.5.}
\end{figure}

 A more precise test is to compare the best-fit to $n$ and
  $Q_{rms-Ps}$ using the summed tensor and scalar contributions
   to CMB anisotropy and comparing to the result if the tensor
    contribution is ignored.  For this test, we have scanned
      the published COBE DMR autocorrelation for
       the  $53A+B \times 90A+B$ channels (Smoot \etal 1992)
	and best-fit to the digitized image, including the
	 cosmic variance. Here, we have used the  fact that the
	  tensor and scalar contributions to the multipole moments
	   add in quadrature, $\VEV{a_{\ell}^2}=\VEV{a_{\ell}^2}_S +
	    \VEV{a_{\ell}^2}_T$ and, for the purpose
	    of this exercise, we have taken the statistical
	    and systematic errors to be
	    uncorrelated.  For $n=0.75,\; 0.5, \;
	     0.0$, we find the difference in $\chi^2$
	     after including the tensor contribution to be $0.5, 0.3, 0.2$,
respectively, for 68 degrees of freedom, a negligible change even
for large tilts.

To distinguish tensor from scalar contributions to the CMB
anisotropy, the COBE DMR observations have to be combined with
measurements on smaller-angular scales.  In Figure 2, we plot
the ratio $T/S$ vs. $\ell$, which shows that the scalar mode
grows relative to  tensor for higher $\ell$'s. Measurements on
smaller-angular scales are sensitive to higher $\ell$
contributions.  Here we have
computed only the Sachs-Wolfe effect; the effect becomes even
more dramatic  for  $\ell\gg 30$,  which is detected
by experiments at 1~degree and
smaller.

\begin{figure}
\vspace*{3.5in}
\caption{The ratio of tensor to scalar fluctuation multipole
moments versus $\ell$.  Note that the ratio decreases for
large $\ell$, as does the net contribution.  Consequently,
the predicted anisotropy at high $\ell$ or, equivalently,
small (1 degree) scales is reduced compared to a scale-invariant
spectrum of scalar fluctuations.}
\end{figure}

 The tensor mode can seriously affect the interpretation of CMB
 measurements for large-scale structure, regardless of the form
 of dark matter.   As an example,  the
 best-fit cold dark-matter (CDM) model  to the COBE DMR
 results assuming $T/S \ll 1$  has  a  bias factor $b\simeq 1$.
 The bias  is a  ratio of the fluctuations in  galaxies
 to the fluctuations in cold dark matter.  The preferred value
 is in the range between 1.4 and 2.
 (Formally, the bias factor $b \equiv 1/\sigma_8$, where
 $\sigma_8$ is the {\it rms} mass fluctuation on
 the scale $8h^{-1}\,$Mpc.)
 If, however, the tensor contribution to the CMB quadrupole
 is significant, then the extrapolated density perturbation
 amplitude at $8h^{-1}\,$Mpc  is reduced, and
  the best-fit CDM model has $b>1$; see Fig.~2.
  Two related effects combine to increase $b$:
  the power spectrum
  is tilted (less power on small scales for fixed quadrupole
  anisotropy), and scalar perturbations only
  account for a fraction of the
  quadrupole anisotropy.  We find, very roughly,
  \begin{equation}  \label{bias}
  b \simeq 100^{(1-n)/2}\sqrt{1+T/S} \simeq 10^{(T/S)/7}\sqrt{1+T/S},
  \end{equation}
  where ``100'' is the ratio of the scale relevant to
  the quadrupole anisotropy, $\lambda \sim 1000h^{-1}\,$Mpc,
  to the scale $8h^{-1}\,$Mpc.  For $T/S = 0.53, 1.4, 2.5,$
  and 3.3, the bias factor $b=1.4, 2.4, 4.6,$ and 7.8
  (and $n = 0.92, 0.78, 0.59$ and 0.44).
  While these numbers should only be taken as rough estimates, the trend
  is clear:  larger $T/S$ permits larger bias.
  [Note that  $b>2$ is probably unacceptable for explaining large
  scale bulk flows and galaxy formation, which suggests a limit
  $T/S<2$.]

  Finally,  the consideration of tensor mode contributions
  greatly affects the expectation for smaller-angular scales.
  Extrapolated to small-angular scales,  the tensor mode
  contribution decreases relative to the scalar contribution
  both in the Sachs-Wolfe effect (see Fig.~2) and even more
  dramatically in the Doppler effect at angles  less than
  a few degrees.
  Consequently, at 1 degree, only the scalar contribution
  should be observed.
  This is a highly relevant point at the present time since
  preliminary measurements  of the CMB anisotropy at 1 degree
  scales (Gaier \etal 1992) indicate a
  low value compared to the extrapolation from COBE DMR
  assuming  a scale-invariant scalar fluctuation spectrum. If the
  result should hold true, then this may be evidence that
   a large fraction of COBE DMR's signal is due to tensor modes.

An alternate explanation is that late re-ionization has washed
out the small-angle fluctuations.  A potential method
  for distinguishing tensor mode from re-ionization scenarios
  is to measure the  ratio of the polarization to the anisotropy.
   Because fluctuations grow after decoupling (while there is
   no coupling to increase the polarization), the ratio
   of polarization to anisotropy decreases
   the  more distant is the last scattering surface.  Hence,
 the  late re-ionization explanation leads to a  greater ratio  than
 the tensor mode domination scenario
  because the last scattering surface is
  at greater red shift in the latter case.
  Polarization-to-anisotropy
  ratio on large-angular scales due to
  tensor plus scalar modes is only about two percent (Polnarev 1985).

 Should the tensor mode be verified as the culprit in suppressing
 the small-angular measurements, it would be a remarkably fortuitous
 breakthrough.  At one stroke,
  we would have a direct handle on  the key cosmological parameters
  that govern large-scale structure, such as the bias factor $b$
  in CDM models  the power-spectrum index $n$,
  and the microphysical parameters that control inflation.

\section*{Acknowledgements}

Since the {\it Journ\'{e}es Relativistes}, several groups in addition to
Davis \etal (1992) have focussed attention on the tensor mode
contribution from inflationary models.  These include
A. Dolgov and J. Silk;
J.E. Lidsey and P. Coles;  L. Krauss and M. White; D. Salopek;
 F. Lucchin, S. Matarrese, and S. Mollerach; A. R. Liddle and
 D. H. Lyth; T. Souradeep and V. Sahni.

This paper reviews work done in collaboration with  our
three valued colleagues, Rick Davis, Hardy Hodges and Michael Turner.
We wish to thank the organizers of the {\it Journ\'{e}es Relativistes}
for inviting us to participate at this
fortuitous juncture and Willem van Leeuwen for his tireless,
 gracious hospitality.
We  also thank
 D.~Bennett, R.~Holman, E.W. Kolb,  S.-H.~Rhie, and L. Kofman
 for useful advice and discussion.
This research was supported in part
by the DOE at Penn (DOE-EY-76-C-02-3071),
at Berkeley (DOE-AC-03-76SF0098).

\end{document}